\documentclass[useAMS,usenatbib]{mn2e}
\usepackage{graphicx,epsfig,amssymb,layout,verbatim,rotating,calc,mathrsfs,natbib}
\usepackage[sumlimits,intlimits,namelimits]{amsmath}
\usepackage{graphics}
\usepackage{rotate}
\usepackage{txfonts}

\newcommand{\ksk}{km~s$^{-1}$~kpc$^{-1}$ }

\title[NGC1300 response models]{NGC~1300 Dynamics:\\ II. The response models}
\author[C.~Kalapotharakos et al.]
{C.~Kalapotharakos,$^{1}$\thanks{ckalapot@phys.uoa.gr
  (CK); patsis@academyofathens.gr (PAP); pgrosbol@eso.org (PG)}
  P.A.~Patsis,$^{1,2,3}$ and P.~Grosb{\o}l$^{3}$\footnotemark[1]\thanks{Based
    on observations collected at the European Southern
Observatory, Chile: program: ESO 69.A-0021.}\\
$^1$Research Center for Astronomy, Academy of Athens, Soranou Efessiou
    4, GR-115 27, Athens, Greece\\
$^2$ Observatoire Astronomique de Strasbourg, 11 rue de l'Universit\'{e},
    67000 Strasbourg, France\\
$^3$ European Southern Observatory, Karl-Schwarzschild-Str. 2, 85748 Garching,
    Germany }

\date{Accepted ..........Received .............;in original form ..........}

\begin{document}

\maketitle

\label{firstpage}

\begin{abstract}
We study the stellar response in a spectrum of potentials describing
the barred spiral galaxy NGC~1300. These potentials have been
presented in a previous paper and correspond to three different
assumptions as regards the geometry of the galaxy. For each
potential we consider a wide range of $\Omega_p$ pattern speed
values. Our goal is to discover the geometries and the $\Omega_p$
supporting specific morphological features of NGC~1300. For this
purpose we use the method of response models. In order to compare
the images of NGC~1300 with the density maps of our models, we
define a new index which is a generalization of the Hausdorff
distance. This index helps us to find out quantitatively which cases
reproduce specific features of NGC~1300 in an objective way.
Furthermore, we construct alternative models following a
Schwarzschild type technique. By this method we vary the weights of
the various energy levels, and thus the orbital contribution of each
energy, in order to minimize the differences between the response
density and that deduced from the surface density of the galaxy,
under certain assumptions.  We find that the models corresponding to
$\Omega_p\approx16$\ksk and $\Omega_p\approx22$\ksk are able to
reproduce efficiently certain morphological features of NGC~1300,
with each one having its advantages and drawbacks.
\end{abstract}



\section{Introduction}

\begin{figure}
\begin{center}
\includegraphics[width=8.5cm]{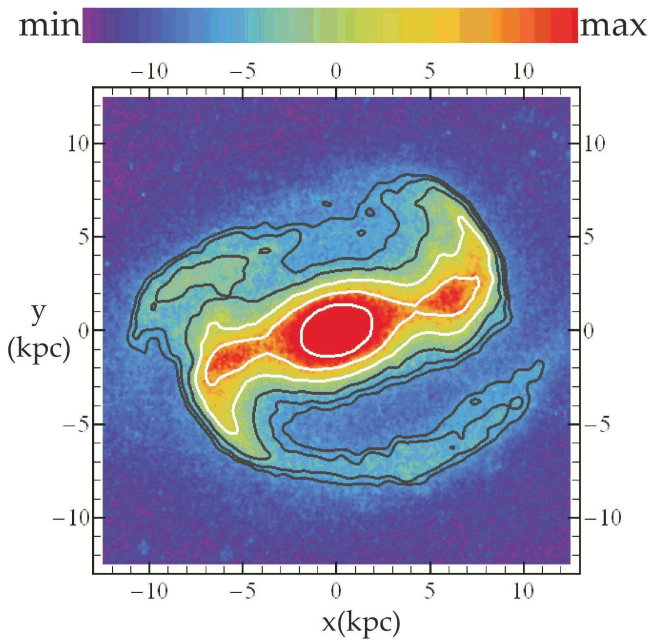}
\end{center}
\caption{The near-infrared surface brightness of NGC~1300,
in gray scale (color scale in online version). This image has been
obtained after applying all the necessary amendments (see Paper I)
and after deprojecting the image with the considered position and
inclination angles (PA, IA)=$(87^{\circ}, 35^{\circ})$. The solid
curves are six iso-density contours describing the basic contour
shapes of NGC~1300. The adopted distance is D=19.6~kpc.}
\label{fig01}
\end{figure}

In \cite{kpg2010i} (hereafter Paper I), we have presented three
different general models representing the potential of NGC~1300. The
morphology we investigate can be observed at a deprojected
near-infrared image of the galaxy (Fig.~\ref{fig01}). Each of these
model-potentials corresponds to different assumptions regarding the
distribution of the luminous matter of the galaxy in the third
dimension, perpendicular to the equatorial plane. They represent
limiting cases for the geometry of the system, and vary from the
pure 2D (Model A) to 3D cases. The two 3D cases correspond either to
a pure cylindrical geometry of a 3D disc with a constant scale
height (Model B) or to a combination of a spheroidal component
representing the central part of the bar with the cylindrical
geometry of a 3D disc for the rest of the luminous mass (Model C).
All of them have the option of the inclusion of two additional terms
in the potential representing the central mass concentration and the
dark halo component respectively. These terms are constrained by the
kinematical data derived by \citet{linetal1997}.

We note that in our calculation we do not take into account
the outer spiral arms that exist beyond the edges of the frame of
Fig.~\ref{fig01}. These spiral extensions are very weak in the
near-infrared, while they are conspicuous in the optical. This
indicates that they consist mainly of young objects, thus their
contribution to the mass distribution of the galaxy is small
(Grosb{\o}l et al. in preparation).

In the present paper we investigate the detailed dynamics in the
three cases using response models (hereafter RM).
As we have already mentioned in Paper I the global dynamical
behavior in rotating galaxies, is crucially determined by the
assumed value of the pattern speed $\Omega_p$. Thus, in this paper
we use the method of response models \citep{patsis2006} in order to
find which $\Omega_p$ values, in each Model (A, B and C), are able
to reproduce the various morphological features of NGC~1300. We
construct our models under the assumption of a single pattern speed
and we will assess the results at the end. For this reason, we
introduce an index that is a generalization of the Hausdorff
distance \citep[see e.g.][]{dd09} so that we can quantify the
(dis)similarity between the density maps of the response models and
the K-band deprojected NGC~1300 image. Comparisons that involve the
surface density of the galactic disc are done under the assumption
of a constant $M/L$ ratio (see Paper I). We remind also, that from
the image we used (Fig.~\ref{fig01}) we have removed point-like
sources, that are likely to be young stellar clusters (Paper I).

This study allows also the comparison between various response
models and consequently indicates, which kind of geometry applies
better to the NGC~1300 case. Nevertheless, our ultimate goal is the
detection of the dynamical mechanisms behind the observed structures
(bar and spiral arms). This needs a detailed analysis of the orbital
stellar dynamics for each case and is presented in \citet{pkg10}
(hereafter Paper III).

Our paper is structured as follows: in Section \ref{s_rm} we present
the method of response models and the corresponding results. In
Section \ref{s_qc} we describe the generalized Hausdorff distance
that helps us quantify the similarity between the various response
models and the galaxy. In Section \ref{s_stmm} we check the ability
of the response models to describe better the morphological features
of NGC~1300 by varying the relative contribution of the energy
levels. Finally, we discuss our conclusions in Section \ref{s_dc}.

\begin{figure}
\begin{center}
\includegraphics[width=8.2cm]{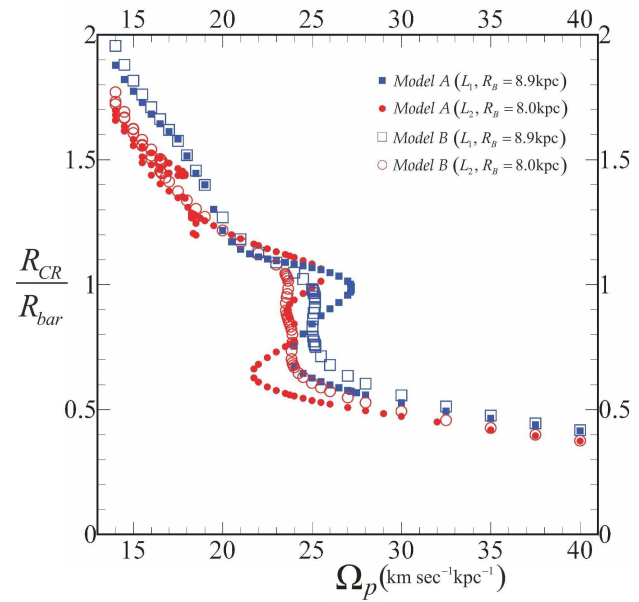}
\end{center}
\caption{The $R_{CR}/R_{bar}$ ratio as a function of the
pattern speed for our models A (filled symbols) and B (empty
symbols). Squares (blue colored in the online version) and circles
(red colored in the online version) correspond to the long and the
short semi-major axis of the bar, respectively. Note that there are
ranges of $\Omega_p$ values that give multiple Lagrangian points.}
\label{figrlrb}
\end{figure}

\section{Response models}\label{s_rm}
These models show the response of an initially axisymmetric stellar
disc, with particles moving at the beginning of the simulation in
circular orbits determined in the axisymmetric part of the
potential.

The steps we follow in our numerical experiments are:
\begin{enumerate}
    \item we choose a potential (among Models A, B, C) and the pattern speed
    value $\Omega_p$,

    \item we populate uniformly the disc up to $R_{max}=15$kpc  (the $(x_i,y_i)$ of $10^6$ particle positions on the plane are taken at random positions),

    \item we set each particle in circular motion (in the rotating frame) with
    velocity
    $v_{circ}=\left(\sqrt{\left|\dfrac{d\Phi_0}{dR}\right|\dfrac{1}{R}}-\Omega_p\right)R$,
    where $\Phi_0$ is the axisymmetric part of the potential,

    \item we grow the non-axisymmetric terms of the potential linearly from
    0 towards their full amplitude within two pattern rotations,

    \item we integrate the particles' orbits for 25 pattern
    rotations (we consider that a particle has escaped and we stop the integration of its orbit when it reaches a
    radius $R>22$~kpc).

    \item Finally we construct density maps, by converting our data files to images. For this we consider a grid and we take into account the numerical density of the test particles on the disc.
\end{enumerate}

\begin{figure*}
\begin{center}
\includegraphics[width=12.0cm]{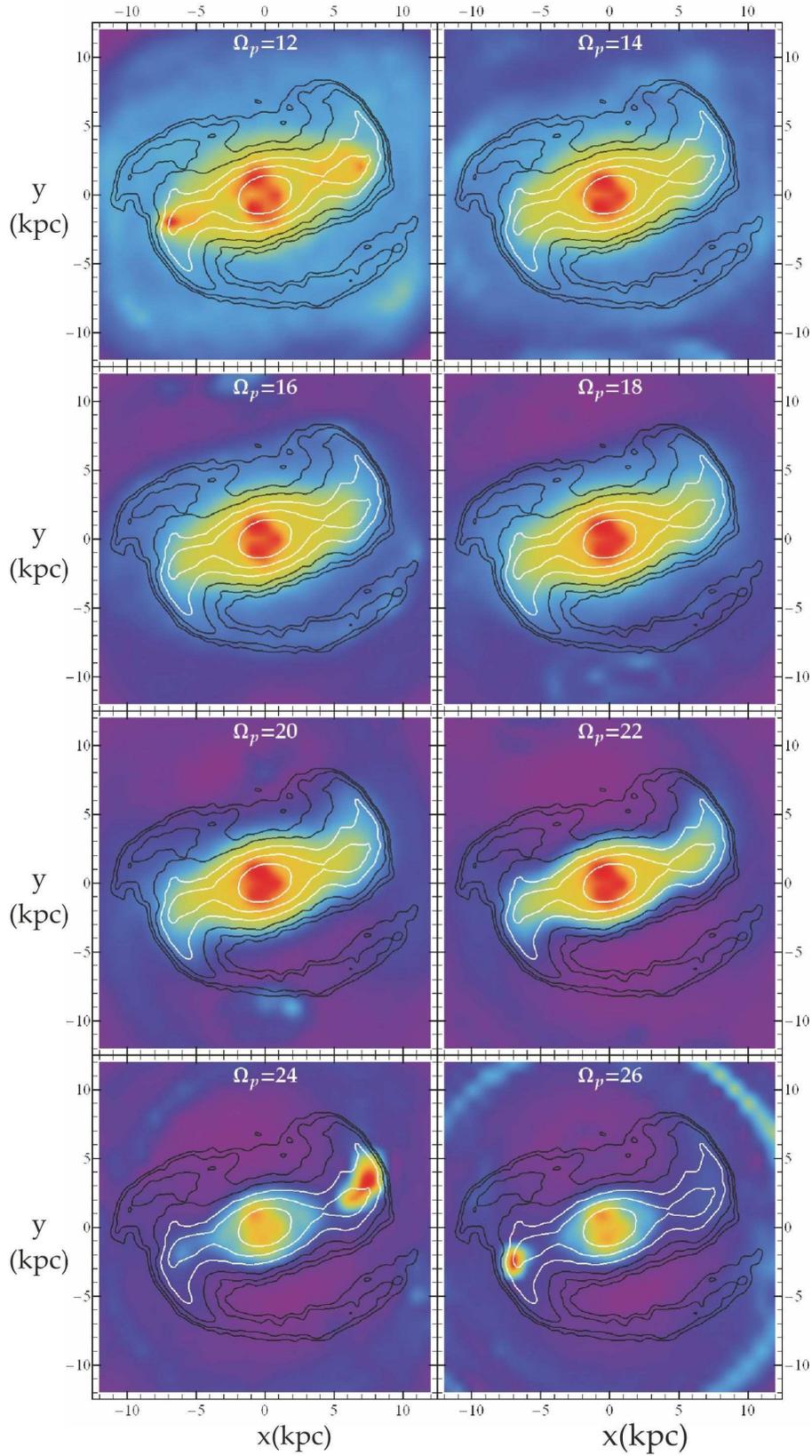}
\end{center}
\caption{The surface density in gray scale (color scale in
the online version) of the RMs corresponding to the potential of the
pure 2D case (Model A). At the top of each panel is indicated the
$\Omega_p$ value. In each frame we plot over the density maps of the
RMs six representative contours (solid lines) of the K-band image of
NGC~1300 (the same with those plotted in Fig.~\ref{fig01}). The
density scale is the same as in Fig.~\ref{fig01}.}
\label{fig02}
\end{figure*}

\begin{figure*}
\begin{center}
\includegraphics[width=12.0cm]{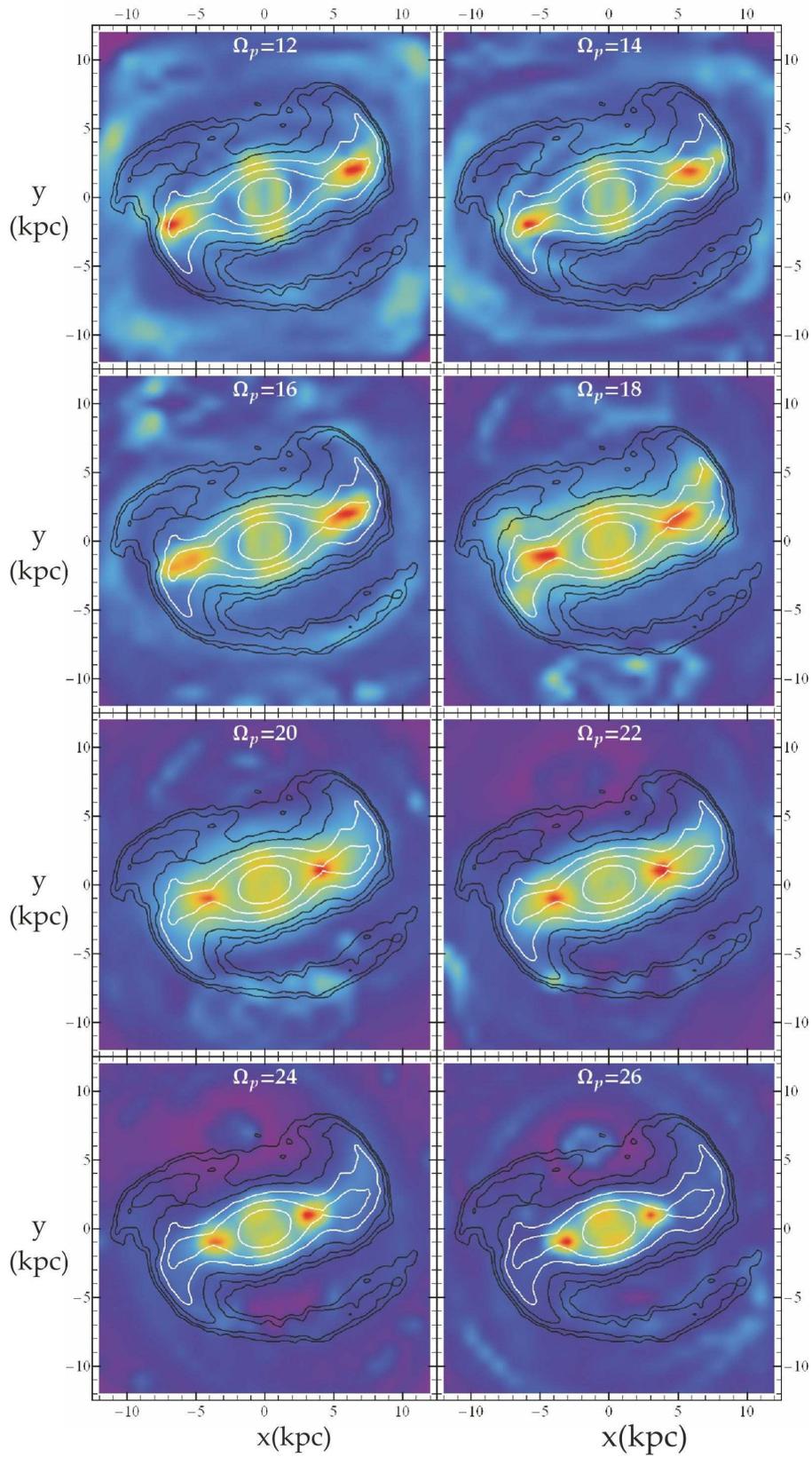}
\end{center}
\caption{As Fig.~\ref{fig02}, but for the potential corresponding to
the cylindrical geometry (Model B).} \label{fig03}
\end{figure*}

\begin{figure*}
\begin{center}
\includegraphics[width=12.0cm]{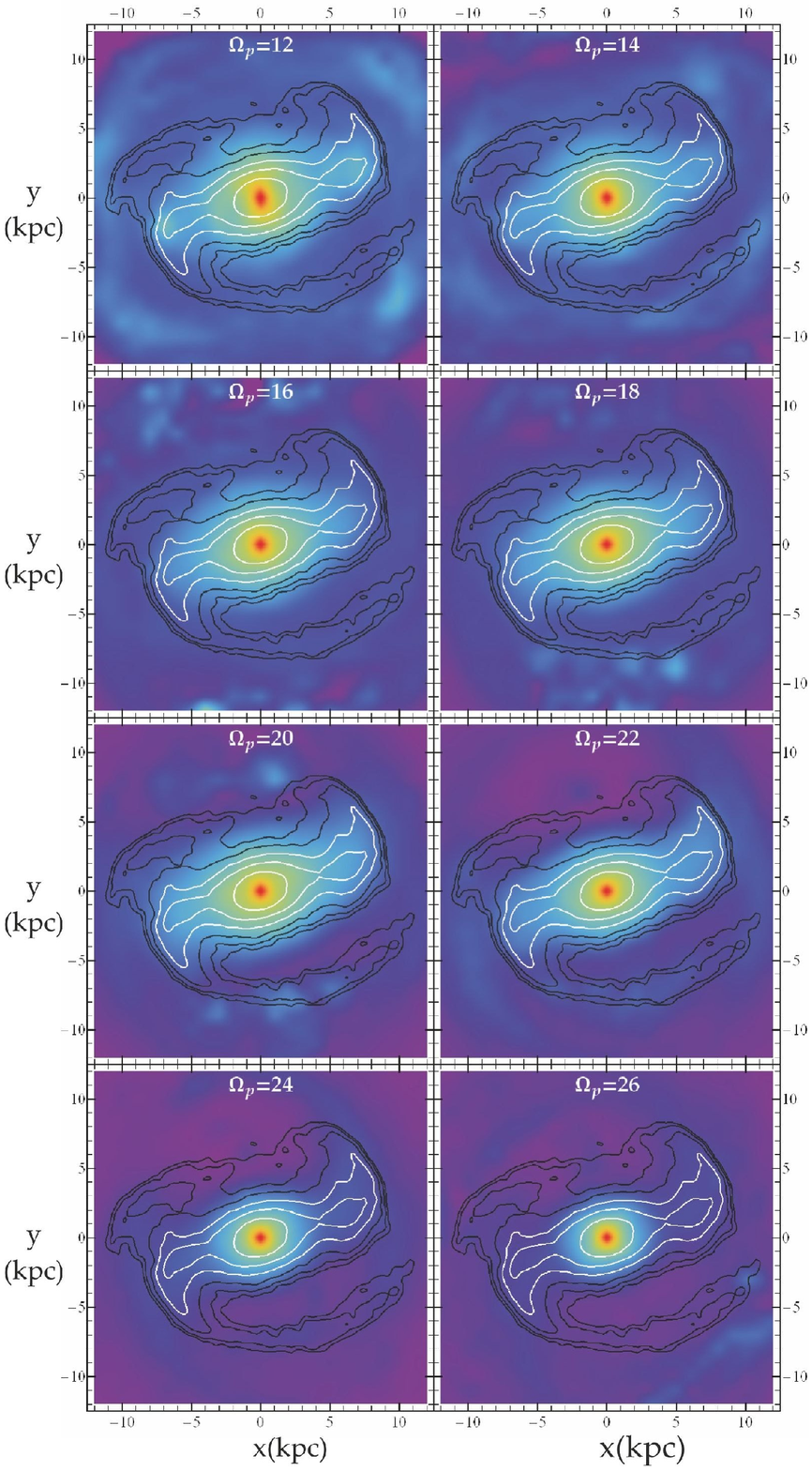}
\end{center}
\caption{As Fig.~\ref{fig02}, but for the potential corresponding to
the (spheroidal + cylindrical) geometry (Model C).} \label{fig04}
\end{figure*}

In Paper I we have studied the distribution of the Lagrangian points
in all general models (A, B, C) and for an extended range of
$\Omega_p$ values (see fig.~10 of Paper I). In
Fig.~\ref{figrlrb} we plot the ratio $R_{CR}/R_{bar}$ of the
corotation radius $R_{CR}$ over the semi-major axis of the bar
$R_{bar}$ as a function of the corresponding $\Omega_p$ value. We
have omitted the points corresponding to Model C since they are very
close to those of Model B. We have considered the bar extending up
to the outer white isocontour of Fig.~\ref{fig01} (third white
contour starting counting from inside). This contour is not
symmetric to the origin, since the right semi-major axis of the bar
is longer than the left one. Thus, we set $R_{bar}=8.9$~kpc for the
right semi-major axis of the bar and $R_{bar}=8.0$~kpc for the left
one. From the above it is obvious that the ratio $R_{CR}/R_{bar}$
depends on the side of the bar. The considered $R_{CR}$ value is the
radius of the Lagrangian points $(L_1, L_2)$ lying in the
corresponding side. In Fig.~\ref{figrlrb} the filled circles and
squares correspond to Model A, while the empty ones to Model B. The
squares (colored blue in the online version) and the circles (colored red in the online version) represent the ratio $R_{CR}/R_{bar}$
on the right and on the left side ($L_1$ and $L_2$ area),
respectively. It is argued that $R_{bar}$ should not be  
longer than the corotation radius $R_{CR}$. This argument is
sustained by dynamical studies according to which the orbital
content beyond corotation is not able to support the bar shape
\citep{cont1980}. In our models, as we have mentioned already in
Paper I, we see that there are ranges of $\Omega_p$ values with
multiple Lagrangian points (especially in Model A). Particularly, in
some cases the ratios $R_{CR}/R_{bar}$ corresponding to multiple
Lagrangian points are quite different when we consider as corotation distance the distance of the one or the other Lagrangian point. E.g. in Model A at
$\Omega\approx25$~\ksk the ratio $R_{CR}/R_{bar}$ varies from
$\approx0.5$ (unphysical value since it is significantly lower than
1) up to $\approx1.1$ (reasonable value). These cases are quite
complicated and lie beyond the standard paradigm. Thus, there is no
a priori knowledge about the orbital behavior and the morphologies
they can support. In any case, the models with
$\Omega_p\gtrsim28$~\ksk have single Lagrangian points corresponding
to significantly low $R_{CR}/R_{bar}$ values.

In this study we present the results corresponding to the $\Omega_p$
values from 12 to $26$~\ksk. Empirically it has been also realized
that beyond the limits of this range the response models were
particularly problematic in the sense that there was obviously no
good agreement with the image of the galaxy. The models with
$\Omega_p\gtrsim17$\ksk correspond to ``fast bars'' as are usually
called these with $R_{CR}/R_{bar}<1.4$ \citep{rsl2008, ds00}.

In Fig.~\ref{fig02} we plot, in gray scale (color scale in
online version), the density maps we obtain from the RMs
corresponding to the potential of Model A (2D case) for the $12 <
\Omega_p <26$~\ksk values. Here, we plot the RMs corresponding to
the 8 indicated $\Omega_p$ values. However, we have data for models with totally 15
$\Omega_p$ values within the above range of $\Omega_p$. On
each panel of this figure we have also overplotted the same (as in
Fig.~\ref{fig01}) six iso-density contours (solid lines) of the
K-band image of NGC~1300, outlining its major morphological
features.

In this figure we observe that the low $\Omega_p$ values (first row)
fail to reproduce both the bar and the spiral structure. The
response bar is conspicuously broader than the galactic bar, while
the spiral arms are absent or flocculent. For higher values of
$\Omega_p$ we get spiral arms in good agreement with the real ones
(see panel for $\Omega_p=16$~\ksk) although the bar is still broad.
For even higher $\Omega_p$ values (third row) the formed spiral arms
of the models  do not match with the spirals of the galaxy.
Nevertheless the bar clearly gets an ansae character in agreement
with the NGC~1300 bar morphology (see panel for $\Omega_p=22$~\ksk
as best example). Finally, the highest $\Omega_p$ values (last row)
produce very open spiral arms and/or rings and a disintegrated bar.

Figure \ref{fig03} presents the RMs for Model B (thick disc,
cylindrical geometry) at the same $\Omega_p$ values as for Model A
(Fig.~\ref{fig02}). In this case the response bars have always high
density regions at their ends, underlying a clear ansae-type
character. A common feature is also a density enhancement along the
minor axis. The bar morphology has a good relation with the galactic
bar for low $\Omega_p$ values (two first rows). This indicates that
a proper selection of the bar-supporting orbits  might lead to a
best fitting of the NGC~1300 bar.  As $\Omega_p$ increases (panels
in the third and forth rows) the response bar becomes shorter than
that of the galaxy. The response spiral arms are not always well
described  but there is a range of $\Omega_p=16-22$~\ksk, where
occasionally we find high densities in the models at the regions
corresponding to the galactic spiral arms.

Figure \ref{fig04} is similar to Figs.~\ref{fig02} and \ref{fig03}
but now for Model C (spherical and cylindrical geometry). The
morphology of the response bar resembles more the morphology we get
in Model A especially for the lower $\Omega_p$ values (first two
rows). However, the inner structure of the NGC~1300 bar ($R\lesssim
2$kpc) is fitted clearly better in this case.  The general spiral
response is closer to that of Model B while it is in general
fainter.

Up to this point, the similarity between the various RMs and the
morphological features of the galaxy have been made by eye. Although
the eye is usually a good selecting tool it lacks of objectivity and
capability of quantification. For this reason, in the next section,
we introduce an index in order to quantify the comparison between
the NGC~1300 morphology and the response models.

\begin{figure*}
\begin{center}
\includegraphics[width=14cm]{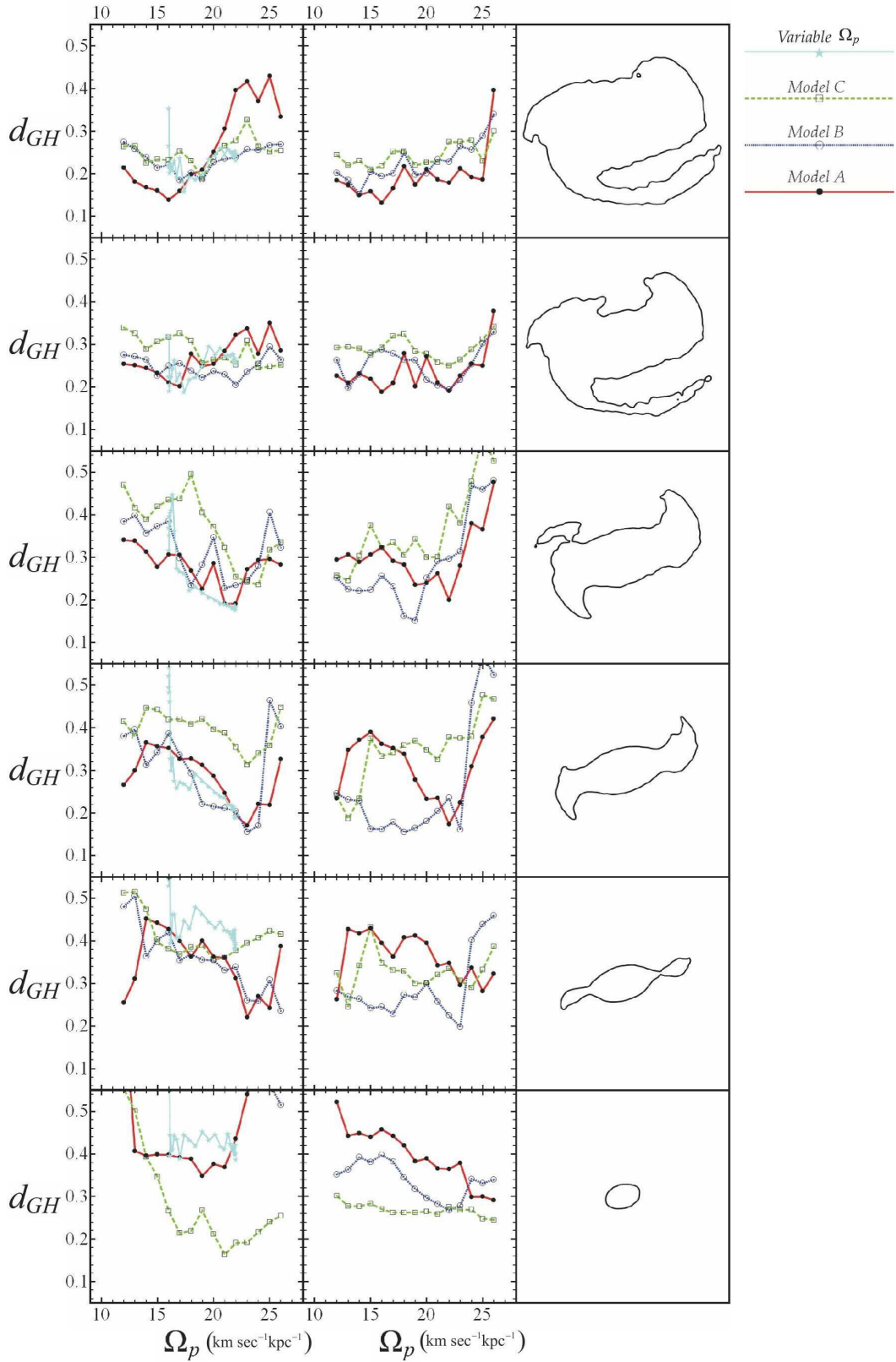}
\end{center}
\caption{In the left-hand column we plot for the RM, the minimum
generalized Hausdorff distance $d_{GH}$ value relative to the
NGC~1300 contour line shown in the right-hand column (same row). In
the middle column we plot the minimum $d_{GH}$ values corresponding
to SMs (see Sect.~\ref{s_stmm}). Note that filled circles,
empty circles and empty squares (red, blue and green colored lines
in the online version) correspond to pure 2D geometry (Model A), to
cylindrical geometry (Model B) and to spheroidal+cylindrical
geometry (Model C), respectively. The star symbol (light blue line
in the online version) in the left-hand column corresponds to a
scenario with time-dependent $\Omega_p$ (see Sect.~\ref{s_dc}).}
\label{fig06}
\end{figure*}

\begin{figure*}
\begin{center}
\includegraphics[width=14.0cm]{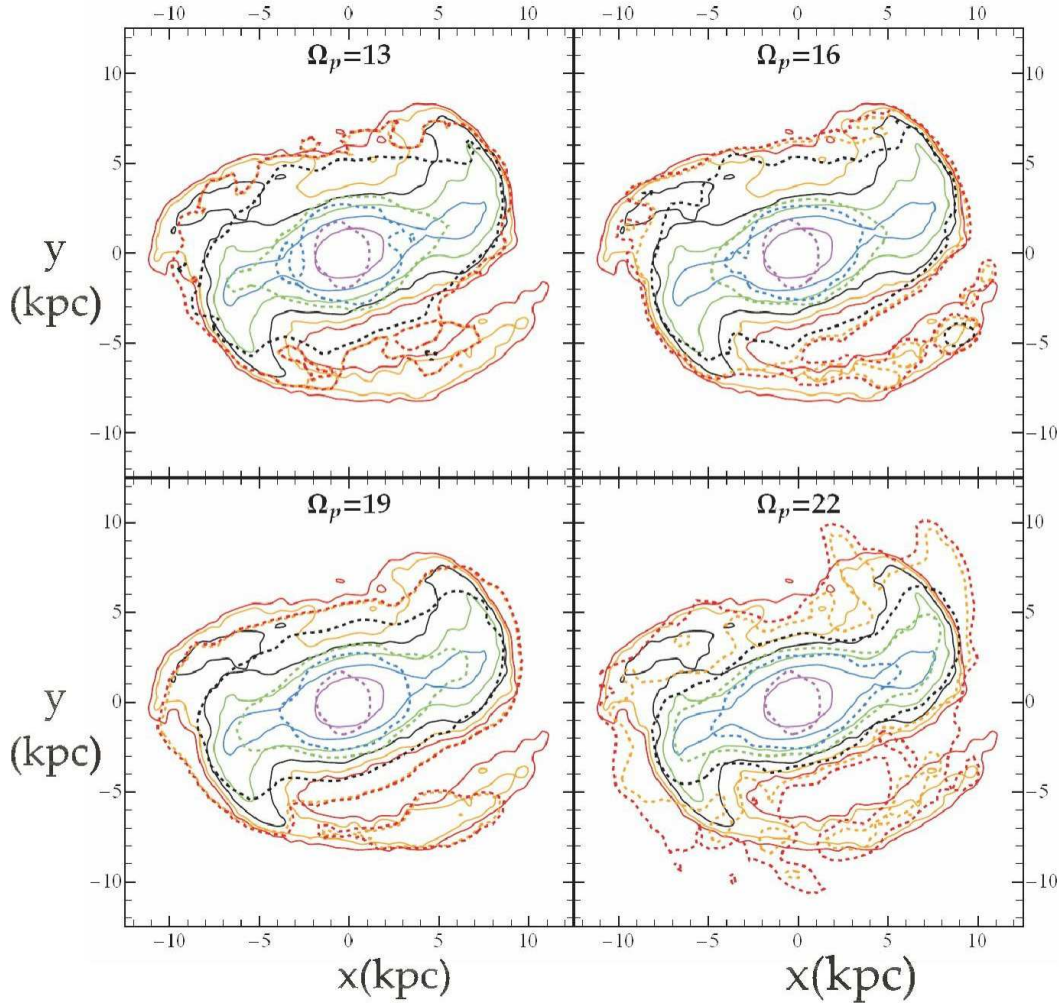}
\end{center}
\caption{The six representative contour lines of NGC~1300 (solid
lines) together with the six contour lines corresponding to the
minimum $d_{GH}$ values (dashed lines) for all the SMs of Model A.
The comparable lines have the same color. There are cases that these
models are better than RMs in reproducing some morphological
features of NGC~1300.} \label{fig10}
\end{figure*}

\begin{figure*}
\begin{center}
\includegraphics[width=14.0cm]{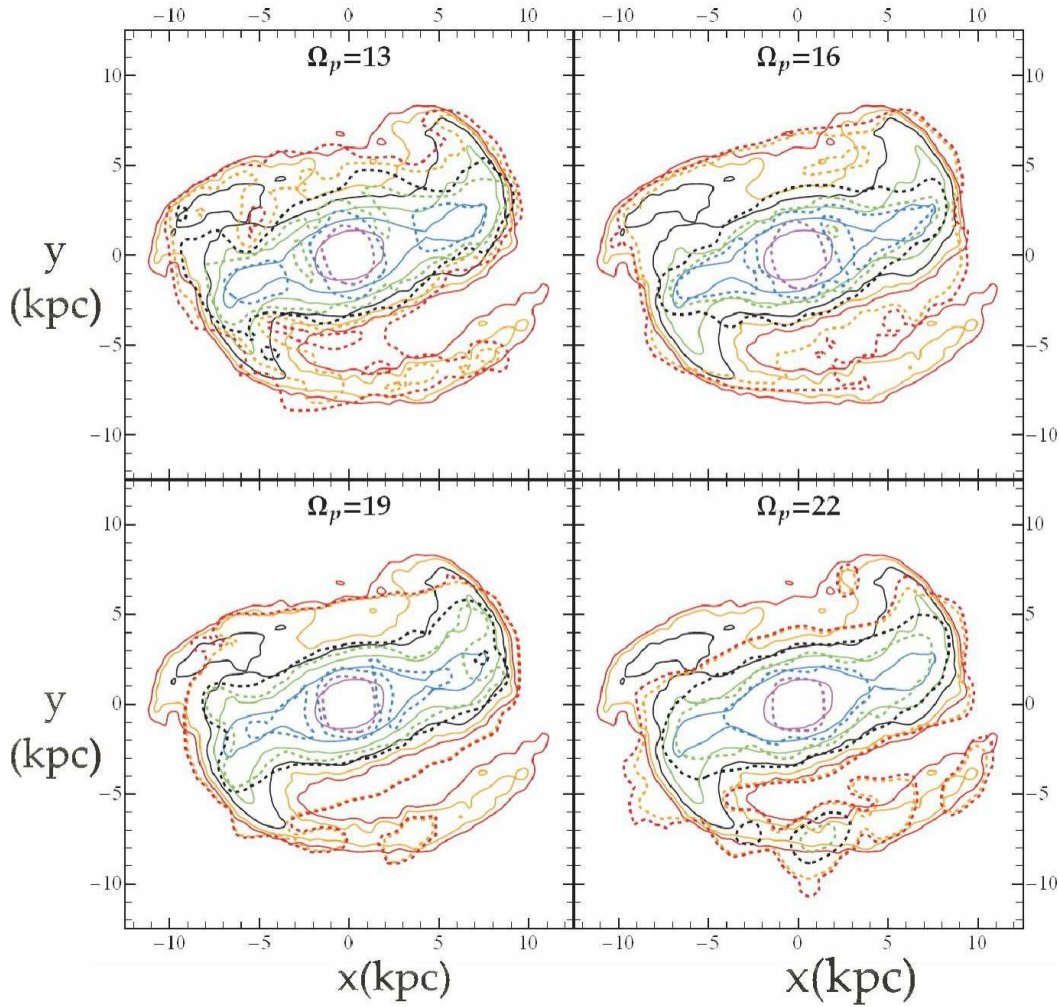}
\end{center}
\caption{Similar to Fig.~\ref{fig10} but for the potential
corresponding to the cylindrical geometry (Model B).} \label{fig11}
\end{figure*}

\begin{figure*}
\begin{center}
\includegraphics[width=14.0cm]{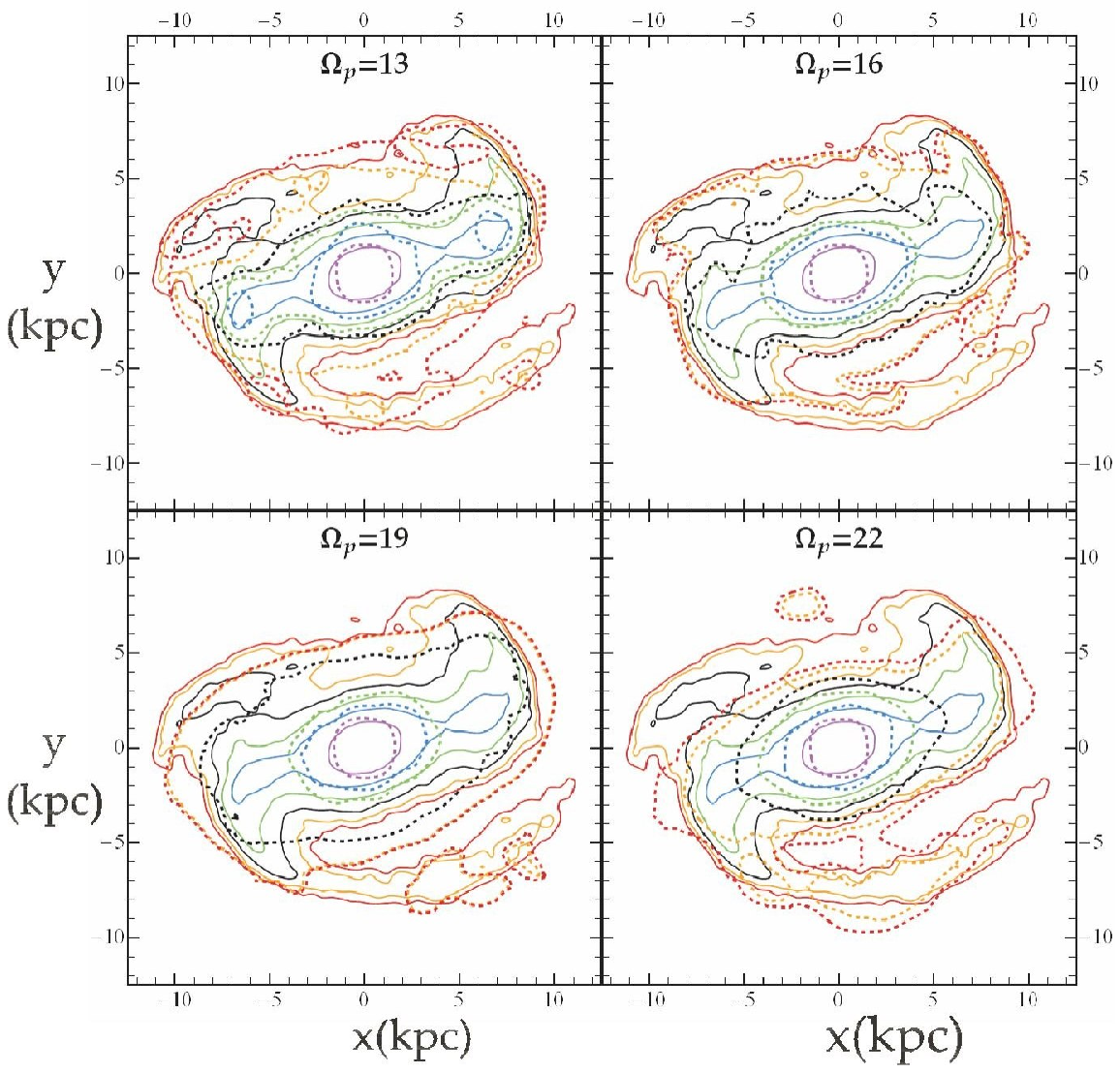}
\end{center}
\caption{Similar to Fig.~\ref{fig10} but for the potential
corresponding to the (spheroidal + cylindrical) geometry (Model C).}
\label{fig12}
\end{figure*}

\section{Quantitative comparison}\label{s_qc}

In Fig.~\ref{fig01} we have plotted six iso-density contours of
NGC~1300 reproducing its basic morphological features. Our goal is
to ``measure'' in an objective way the similarity of an RM density
map with the galactic morphology. Thus, we want to establish a
quantitative criterion for this resemblance.

The methodology of this study is related with image comparison and
pattern recognition techniques. A simple and well known index used
in such studies is the ``Hausdorff Distance'' $d_{H}$, that is
defined as follows: let $\mathbf{A}$, $\mathbf{B}$ be two non-empty
subsets of a metric space $(\mathbf{M},d)$. In this case the
Hausdorff distance $d_{H}$ between $\mathbf{A}$ and $\mathbf{B}$
reads
\begin{equation}\label{hd}
    d_{H}=\max\left(\sup_{a\in \mathbf{A}}\inf_{b\in \mathbf{B}}d(a,b),~\sup_{b\in \mathbf{B}}\inf_{a\in
    \mathbf{A}}d(a,b)\right).
\end{equation}
Considering two sets of points corresponding to two contour lines
and $d$ being the Euclidean distance, $d_{H}$ measures the maximum
of the distances between the points of each curve from the other one
(as a whole). It is obvious from the above definition that $d_{H}=0$
corresponds to identical lines. The major issue about $d_{H}$ is its
sensitivity to outliers. Let us suppose that a contour line consists
of two parts, the major one being identical to a contour line of
NGC~1300, while the minor one being a tiny part far away. In such a
case although we have a good matching of the response model with the
galaxy morphology (at least for the specific contour), $d_{H}$ reads
a false high value. There have been proposed many modifications of
$d_{H}$ that aim to reduce this drawback \citep[][and references
therein]{zsd2005}. Here we introduce an index which is a
generalization of $d_{H}$ and it is appropriate for our study.

We call this index ``Generalized Hausdorff Distance'', $d_{GH}$, and
it is defined as follows: let $\mathbf{C_1}$, $\mathbf{C_2}$ be two
curves on the plane of non-zero length $L_{C_1}$ and $L_{C_2}$,
respectively. The $d_{GH}$ between $\mathbf{C_1}$ and $\mathbf{C_2}$
is
\begin{equation}\label{ghd}
    d_{GH}=\frac{1}{L_{C_1}}\int_{b\in C_1}\frac{\inf\limits_{a\in C_2}
    d(a,b)}{\|b\|}dl+\frac{1}{L_{C_2}}\int_{b\in C_2}\frac{\inf\limits_{a\in C_1}
    d(a,b)}{\|b\|}dl,
\end{equation}
where the integrations are over the curves $C_1$ and $C_2$ and
$\|\cdot\|$ denotes the Euclidean norm. The integration adds the
contributions from all points, while the division with the
curve lengths and the norm $\|b\|$ aims to the normalization of
the final $d_{GH}$ value so that it can be comparable for two
totally different contour lines. In practice, for the computation
of $d_{GH}$ we sample both curves with a considerable number
of points $N_{C_1}$, $N_{C_2}$ and we calculate the quantity
\begin{equation}\label{ghdp}
    d_{GH}=\frac{1}{N_{C_1}}\sum_{i=1}^{N_{C_1}}\frac{1}{{\|b\|}}\min\limits_{\substack{a_i\in
C_1\\b\in
    C_2}}d(a_i,b)+\frac{1}{N_{C_2}}\sum_{i=1}^{N_{C_2}}\frac{1}{{\|b\|}}\min\limits_{\substack{a_i\in
C_2\\b\in C_1}}d(a_i,b).
\end{equation}
What we do is to find which contour line of each RM corresponds to
the minimum $d_{GH}$ value relative to each one of the contour lines
of NGC~1300 we have plotted in
Figs.~\ref{fig01},\ref{fig02}-\ref{fig04}.

In Fig.~\ref{fig06} (left-hand column) we have plotted the minimum
$d_{GH}$ values corresponding to the NGC~1300 contour lines shown in
the right-hand column of the same figure. Filled circles, empty
circles and empty squares (red, blue and green colored lines in the
online version) correspond to Models A, B and C, respectively.
Note that in this figure we have also included the models
corresponding to $\Omega_p$ values not plotted in
Figs.~\ref{fig02}-\ref{fig04}. Based on this criterion, the basic
trends that are concluded by observing the behavior of $d_{GH}$ for
all the Models (A, B, C) and for the studied range of $\Omega_{p}$
values are:
\begin{itemize}
    \item the lower $\Omega_p$ values in Model A are more
    efficient (than the higher ones) in portraying the outer contours corresponding mainly to the
    spiral structure of NGC~1300 (see the two first rows of
    Fig.~\ref{fig06})
    \item the higher $\Omega_p$ values in Model A are more
    efficient (than the lower ones) in portraying the inner contours corresponding to the
    barred structure of NGC~1300 (see third to fifth rows of Fig.~\ref{fig06})
    \item Model B is better than Model A in portraying the spiral
    structure in high $\Omega_p$ values, while for low  $\Omega_p$'s, Model A is in general better than B, especially at the outermost isocontour (first row of Fig.~\ref{fig06}).
    \item Model C is in general worst than Models A and B in
    portraying the spiral and bar structure, while it is definitely
    superior in portraying the very inner area especially for the
    middle and high $\Omega_p$ values (see last row of
    Fig.~\ref{fig06}).
\end{itemize}
Nevertheless, we should have in mind, that the quantification of the
similarity of the contours of the model with those of the galactic
image by means of $d_{GH}$, cannot in general account for small
morphological differences, essentially in the phases, of the two
curves. Despite this drawback $d_{GH}$ is a reliable index in most
cases encountered in the present study.

All the above results are related with the response models and
depend to a large extent on their initial set up. These models
provide useful information about the underlying dynamics and the
morphological features they can support. Despite the fact that the
setting up of the models is done in an unbiased way, there is no
physical reason to exclude weighting some energies more than others
in an effort to improve the similarity between RMs and galaxy
morphology. In order to explore this we modify our models by
introducing weights attached to each energy level. The energies of
the particles are determined by giving them circular velocities in
the axisymmetric potential. This determines where they will be end
up in the well of the full potential. However, one can consider a
totally different energy distribution of the particles of the model.
The procedure and the corresponding results are described in the
following section.

\section{Weighting energy levels}\label{s_stmm}

\citet{schw79} in a pioneer study, presented a method for
constructing time independent self-consistent solutions of specific
models. His method has been widely used since then
\citep{rich80,rich82,schw82,rich84,rictre1984,schw93,mfr96,cdzmr99,acdm2010}
mostly for models representing elliptical galaxies. The basic idea
of this method is very simple:
One considers a density distribution function and then finds the
associated potential via the Poisson equation. The next step is to
calculate a library of orbits corresponding to this potential for
the desired energy range. By assigning a weight to each orbit one
tries with the superposition of the weighted orbits to reproduce the
initially imposed density.
%
%
%
For models of normal spiral galaxies, a similar method has been used
in the past by \citet{pcg91} for the investigation of the
self-consistency in 12 galaxies, while in barred-spiral systems has
been applied by \citet{kcont96}.

In practice what one is looking for is a way to minimize the
difference between the imposed density and that corresponding to the
superposed orbits. This detailed task is very demanding and is
beyond the scope of this paper. Here we will construct our ``best''
models acting in a more rough way. We follow the rationale of
Schwarzschild's method but we assign weights to the various energy
levels. The density distribution for each energy level is considered
being the one we get from the corresponding RM. This assumption is
true only for phase space domains where Chaos dominates and for time
scales for which the corresponding initial conditions have been
evolved and are close to a dynamical equilibrium due to chaotic
mixing. For phase space domains (of specific energy levels) with
significant fractions of regular orbits the above assumption is just
a simplification since the distribution of the particles on the
existing tori can be supposed to be different than that
corresponding to the RMs.

The determination of the weights is made by the following procedure.
Let $N_e$ be the number of Jacobi constant (energy)
intervals, which we give to the particles of a response model and $N_c$ the
number of grid cells we divide the plane $(x,y)$. In this situation we
derive the weights $w_i, (i=1,\ldots,N_e,~w(i)\geq0),$ that minimize
the quantity
\begin{equation}\label{minim}
    \sum_{j=1}^{N_c}\left(\sum_{i=1}^{N_e}w_i\frac{m_{e_{ij}}}{m_j}-1\right)^2
\end{equation}
where $m_{e_{ij}}$ is the mass contribution of the $i-$th energy
interval on the $j-$th grid cell and $m_j$ the mass contribution we
consider corresponding to the same region at the NGC~1300 image on
the $j-$th grid cell. We consider the ratio $m_{e_{ij}}/m_j$ in
order to treat equally  high and low density regions. The final
considered surface density distribution is given by
\begin{equation}\label{bestden}
    d_j=\frac{\sum_{i=1}^{N_{e}}w_i
    m_{e_{ij}}}{S_j},~~~~j=1,\ldots,N_c
\end{equation}
where $d_j,~S_j$ are the density and the surface area of the $j-$th
grid cell, respectively. In the present study, we used $N_e=30$ and
$N_c=24\times24=576$ for the central 576~kpc$^2$ area.

Thereupon, we repeat the procedure of the determination of the
contour lines of each of these ``Schwarzschild type'' models
(hereafter SM) corresponding to the minimum $d_{GH}$ value relative
to each contour line of NGC~1300 plotted in Figs.~\ref{fig01} and
\ref{fig02}-\ref{fig04}.

\begin{figure*}
\begin{center}
\includegraphics[width=13cm]{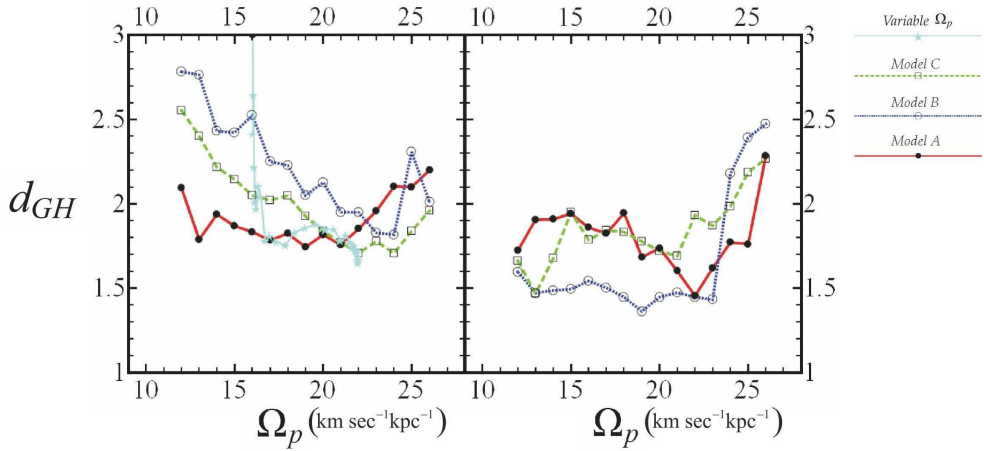}
\end{center}
\caption{The ``global'' $d_{GH}$ index for all the six contour lines
presented in Fig.~\ref{fig06}. The left  panel corresponds to RMs,
while the right one to SMs. These  $d_{GH}$ values represent
globally a model. The plotting styles are as in
Fig.~\ref{fig06}.} \label{fig05}
\end{figure*}

Going back to Fig.~\ref{fig06}, in the middle column, we plot the
minimum $d_{GH}$ values, similar to the panels in the left-hand
column (RMs), but for the density distribution corresponding to
Eq.~\eqref{bestden} (SMs). The comparison between the RMs and SMs
behavior of $d_{GH}$,  shows the following:
\begin{itemize}
    \item In 2D geometry (Model A, filled circles or red color
    line in the online version) the SMs with the high
    $\Omega_p$ values are more efficient (than the corresponding RMs)
    in portraying the spiral structure of NGC~1300 (panels in the two first rows in Fig.~\ref{fig06}).

    \item In cylindrical geometry (Model B, empty circles or blue
    color line in the online version) the SMs with the low
    $\Omega_p$ values are more efficient (than the corresponding RMs)
    in portraying the bar of NGC~1300 (third to sixth rows of Fig.~\ref{fig06}).

    \item In spheroidal+cylindrical geometry (Model C, empty squares
    or green color line in the online version) the SMs
    with the lowest $\Omega_p$ values ($\Omega_p<16$~km~sec$^{-1}$kpc$^{-1}$) are more efficient
    (than the corresponding RMs) in portraying the NGC~1300 bar
    (third to sixth rows of Fig.~\ref{fig06}).
\end{itemize}

In Figs.~\ref{fig10}-\ref{fig12} we plot the isocontours
(dashed lines) corresponding to the minimum $d_{GH}$ with respect to
the isocontours of NGC~1300 (solid lines), for a sample of the SMs
of the Models A, B and C, respectively. These samples include
characteristic cases such as those with $\Omega_p=16$~\ksk and
$\Omega_p=22$~\ksk. Note that the corresponding isocontour curves (of
model and galaxy) are plotted in the same color. From these
figures, it becomes evident the nice behavior of the $d_{GH}$ index
in quantifying the comparison between the models and the image of the
galaxy. In almost all cases low (or high) $d_{GH}$ values imply good
(or bad) resemblance between the corresponding curves.

Since the $d_{GH}$ values for the various isocontours are by
definition comparable, we can add all the values of each model in
order to get a unique value describing globally the model. Such an
index puts of course equally weights to all isocontours and it could
be low valued in models with no matching curves. Nevertheless, it is
reliable in cases with many model curves showing moderate
resemblance with those of the galactic image. In Fig.~\ref{fig05} we
plot, for all RMs and SMs, this ``global'' $d_{GH}$ index for the
isocontours of Fig.~\ref{fig06}.

The general conclusions we draw by observing Fig.~\ref{fig05} can be
summarized as follows:
\begin{itemize}
    \item For $\Omega_p<20$~\ksk values, the RM
    (left-hand panel of Fig.~\ref{fig05})
    with the lowest global  $d_{GH}$ index are those corresponding to
    Model A.

    \item The RMs corresponding to Model B and C improve
    their behaviour from low $\Omega_p$ values towards high
    $\Omega_p$ values (for Model C only up to $\Omega_p=24$~\ksk$\!$.)

    \item The overall behavior of SMs corresponding to Model A are
    improved (relative to those of the corresponding RMs) only for
    $\Omega_p$ values around  $\Omega_p=22$~\ksk$\!$.

    \item The overall behavior of SMs corresponding to Model B are
    improved (relative to those of the corresponding RMs) for
    $\Omega_p\lesssim23$~\ksk$\!$.

    \item The overall behavior of SMs corresponding to Model C are
    improved (relative to those of the corresponding RMs) only for
    $\Omega_p\lesssim14$~\ksk$\!$.
\end{itemize}

\section{Discussion and conclusions}\label{s_dc}
We studied the stellar response, under the assumption of a single
pattern speed, for the potential models of the barred-spiral galaxy
NGC~1300 estimated in Paper I and for a wide range of $\Omega_p$
values. The three potential models of Paper I, correspond to three
different geometries as regards the mass distribution in the third
dimension (perpendicularly to the equatorial plane). Our goal was to
find out, which geometries and which values of $\Omega_p$ were able
to reproduce the fundamental morphological features of NGC~1300,
i.e. the bar and the spirals. We used the method of response models
(RM), as in \citet{patsis2006}. So, we start with a uniformly
populated disk of stars moving initially in circular orbits, which
are determined in the axisymmetric part of the potential. We
integrate the orbits of the test particles for many pattern
rotations under the full potential and we get the density maps of
the response models as described in Sect.~\ref{s_rm}.

The ``by eye''  comparison between the RM morphologies and the image
of the galaxy gives useful qualitative information but it lacks of
objectivity and quantification. For this reason we introduced a new
index which is a generalized modification of the Hausdorff distance.
This index quantifies the resemblance of two isocontour lines and
enables the determination of the best matching contour lines (of a
model) to each one of the six preselected isocontours of the
deprojected K-band image of NGC~1300 (Fig.~\ref{fig01}).
This new index testifies the general remarks made by eye and
furthermore it describes the quality of the fitting for the individual components of the galaxy.
Moreover, we
investigated possible improvements of the response models as regards
their resemblance with the NGC~1300 morphology, by assigning weights
to the contributions of the various energy levels. This method
follows the general concept of Schwarzschild's methods, but it is
simpler in the sense that it requires weights for the energy levels
and not for individual orbits. Some of these modified Schwarzschild
models (SM) improve the similarity of specific features of NGC~1300
in comparison to the original response models.

Below we enumerate our conclusions:
\begin{enumerate}
\renewcommand{\theenumi}{(\arabic{enumi})}
    \item The spiral structure of NGC~1300 is reproduced:
    \begin{itemize}
        \item \textit{Best} by the RM corresponding to the potential
        of the pure 2D geometry and for $\Omega_p$ values
        around 16~\ksk$\!$. The big advantage of this model is that the spirals of RM and galaxy have similar pitch angles. The density enhancement of the the RM at the spiral region is along the same direction as the corresponding arms of NGC~1300.
        \item \textit{Acceptable} by the SM corresponding to the potential
        of the pure 2D geometry and for $\Omega_p$ values
        around 22~\ksk$\!$. However, the drawback of this case is that the spirals, especially the one appearing at the lower part of the figures, have different pitch angles than that of the galaxy. The spiral arms of the model share common regions with the NGC~1300 arms but essentially cross each other. They also add additional features beyond the region within which the NGC~1300 barred-spiral structure extends.
        \item \textit{Moderately} by the RM and SM corresponding to the potential
        of the 3D cylindrical geometry and for $\Omega_p$ values
        around 22~\ksk$\!$. However, the remarks about the pitch angles of the spirals we mention above for the 2D case, yield for the thick disc 3D case as well. On top of that the upper arm is not nicely reproduced by this model.
    \end{itemize}

    \item The ansae-type bar  of NGC~1300 is reproduced:
    \begin{itemize}
        \item \textit{Best} by the RM and SM corresponding to the potential
        of the pure 2D geometry, for $\Omega_p$ values
        around 22~\ksk$\!$ (very good is also the $d_{GH}$ index of the RM corresponding to the potential
        of the 3D cylindrical geometry again for $\Omega_p$
        around 22~\ksk$\!$. However, the ansae are reproduced in the correct regions only in the 2D case)
        \item \textit{Very well} by the SM corresponding to the potential
        of the 3D cylindrical geometry and for $\Omega_p$ values
        around 17~\ksk$\!$.
        \item \textit{Moderately} by the SM corresponding to the potential
        of the 3D cylindrical geometry and for $\Omega_p$ values
        around 23~\ksk$\!$.
    \end{itemize}

    \item The inner oval-like shape structure of NGC~1300 is reproduced:
    \begin{itemize}
        \item \textit{Best} by the RM and SM corresponding to the potential
        of the 3D spheroidal+cylindrical geometry and for almost all
        $\Omega_p$ values
        \item \textit{Well} by the SM corresponding to the potential
        of the 3D cylindrical geometry and for $\Omega_p$ values
        around 22~\ksk$\!$.
        \item \textit{Moderately} by the SM corresponding to the potential
        of the pure 2D geometry and for $\Omega_p$ values
        around 23~\ksk$\!$.
    \end{itemize}
\end{enumerate}

All the above indicate the tendencies of the models to improve their
similarity with the K-band image of the galaxy for a given geometry
or for a certain $\Omega_p$ range. However, the relation between
``good'' models and the NGC~1300 morphology is not unambiguous. It
is obvious that some geometries support better certain morphological
features. This may suggest the appropriate geometry for the
description of specific regions of the galaxy (e.g. the central area
should be considered as a spheroidal). Another issue is that a
morphological feature may be reproduced equally well by models that
differ in their pattern speeds. In addition we remind our assumption
of the time-independency of the potential and a constant
time-independent $\Omega_p$. The study of the stellar dynamics of
our system under these assumptions not only helps us see some basic
trends in the dynamical behaviour that favours the formation of the
one or the other morphological feature, but after all it will
indicate the necessity of different assumptions. The evolution of
the models under a time-dependent potential and $\Omega_p$ is very
dubious in the response model approach, since it requires specific
assumptions for the evolution laws for both potential and
$\Omega_p$. The assumptions are free and therefore the number of the
possible combinations is large. Such a study is beyond the scope of
this paper and should be attacked by means of $N$-body simulations.
However, in order to get a crude impression about how different can
be the resulting morphologies if a basic parameter varies with time,
we test a scenario with time dependent $\Omega_p$ in a fixed
potential. We have chosen for our calculations a pure 2D geometry.
The choice is justified by the fact that the RM corresponding to the
pure 2D geometry portrays on the one hand the spiral structure for
$\Omega_p\approx 16$~\ksk and on the other hand the ansae-type bar
for $\Omega_p\approx 22$~\ksk$\!$. We adopted a rather arbitrary
evolution law for the increase of $\Omega_p$, which reads
\begin{equation}\label{decom}
    \Omega_p(t)=19+3\tanh\left(\frac{1}{5}(t-15)\right)\text{\ksk$\!$,}
\end{equation}
where the time $t$ is measured in periods corresponding to a pattern
rotating with $\Omega_p=23$\ksk$\!$. The integration time of the
orbits was up to $t=30$. In this numerical experiment the initial
value of $\Omega_p = 16$~\ksk, while the final one is
$\approx22$~\ksk. In the left-hand columns of Figs.~\ref{fig06} and
\ref{fig05} we have plotted (with star symbol or light blue lines in
the online version) the $d_{GH}$ values corresponding to each
snapshot's $\Omega_p$ value. In these figures we observe that for
the time corresponding to $\Omega_p\approx17$~\ksk and
$\Omega_p\approx22$~\ksk we get a good resemblance with many
isocontours of NGC~1300.
This could be an indication that the present NGC~1300 structure is
just a snapshot in an evolutionary scenario, according to which the
observed morphology undergoes successive transient phases. However,
whatever good results we find, we find for \textit{increasing }
$\Omega_p$ from one value to the other. The opposite procedure, i.e.
decreasing $\Omega_p$ according to the law in Eq.~\ref{decom} does
not lead to equally nice results. Decrease of $\Omega_p$ during an
$N$-body simulation is frequently observed due to dynamical friction
\citep{chandra43, ds93, des98}, but there is no obvious physical
reason for the opposite. In our model the  particles that
participate in the spiral structure stay for longer times at the
outer regions of the disc, i.e in regions that evolve dynamically
slower than particles at the central region that contribute to the
bar structure. So, by increasing $\Omega_p$, the particles that stay
longer at the outer disc do not have enough time to feel the
potential. On the other hand particles staying at the central region
of the system respond fast during the period our simulation reaches
the fast $\Omega_p$ domain.

Concluding, we remark that taking into account all the models we
constructed as well as the techniques we used to assess them, there
are clearly two $\Omega_p$ intervals, where we find morphological
features of the models resembling the  structures of NGC~1300. These
are $\Omega_p$ values close to 16 and 22~\ksk.

Comparing this result with pattern speeds proposed by other authors
for NGC~1300 \citep{e89b, lk96}, there is a relative agreement
between the fast model of \citet{lk96} and the group of models
around $\Omega_p$=22~\ksk, which we find matching the NGC~1300 bar
(\citet{lk96} propose $\Omega_p$=20~\ksk for a distance of the
galaxy D=20~Mpc). The alternative solution proposed by the same
authors ($\Omega_p$=12~\ksk) is in a range of $\Omega_p$ values,
where the bars of our RMs are much thicker than the one of NGC~1300
and the stellar spirals are absent. As regards the work of
\citet{e89b}, the proposed pattern speed gives equilibrium points in
a narrow zone compatible again with the $\Omega_p$=22~\ksk values,
as it will become evident in Paper III.

In any case, the understanding of the observed NGC~1300 morphology
goes through the understanding of the underlying dynamical
mechanisms and this can be done only by studying the orbital
behavior of the ``successful'' models. This will allow us also the
comparison among them, and is done in Paper III in this series.

\section*{Acknowledgments}
We thank Prof. G.~Contopoulos and Dr. Ch. Efthymiopoulos for
fruitful discussions. P.A.P thanks ESO for a two-months stay in
Garching as visitor, where part of this work has been completed.
This work has been partially supported by the Research Committee of
the Academy of Athens through the project 200/739.

\bibliographystyle{mn2e}
\bibliography{evk}

\label{lastpage}

\end{document}